# PERFORMANCE EVALUATION OF MACHINE LEARNING ALGORITHMS FOR INTRUSION DETECTION SYSTEM

**Sudhanshu Sekhar Tripathy [1] and Bichitrananda Behera [2]**

[1, 2] *Department of Computer Science and Engineering, C.V. Raman Global University Bhubaneswar, Odisha.*
*Email: [1] tripathysudhanshu6@gmail.com, [2] bbehera19@gmail.com*

**Abstract**

The escalation of hazards to safety and hijacking of digital networks are among the strongest perilous difficulties that must be addressed in the present day. Numerous safety procedures were set up to track and recognize any illicit activity on the network's infrastructure. IDS are the best way to resist and recognize intrusions on internet connections and digital technologies. To classify network traffic as normal or anomalous, Machine Learning (ML) classifiers are increasingly utilized. An IDS with machine learning increases the accuracy with which security attacks are detected. This paper focuses on intrusion detection systems (IDSs) analysis using ML techniques. IDSs utilizing ML techniques are efficient and precise at identifying network assaults. In data with large dimensional spaces, however, the efficacy of these systems degrades. Correspondingly, the case is essential to execute a feasible feature removal technique capable of getting rid of characteristics that have little effect on the classification process. In this paper, we analyze the KDD CUP-'99' intrusion detection dataset used for training and validating ML models. Then, we implement ML classifiers such as "Logistic Regression, Decision Tree, K-Nearest Neighbour, Naïve Bayes, Bernoulli Naïve Bayes, Multinomial Naïve Bayes, XG-Boost Classifier, Ada-Boost, Random Forest, SVM, Rocchio classifier, Ridge, Passive-Aggressive classifier, ANN besides Perceptron (PPN), the optimal classifiers are determined by comparing the results of Stochastic Gradient Descent and back-propagation neural networks for IDS", Conventional categorization indicators, such as "accuracy, precision, recall, and the f1-measure", have been used to evaluate the performance of the ML classification algorithms.

## 1. INTRODUCTION

In computer networks, the number of fraudulent operations, intrusions, and attacks has increased dramatically in recent years. Owing to innovations in technology progress, more than 90 percent of practical-circumstances activities are currently accessible in cyberspace. Several procedures involving financial services, buying something, examinations via the Internet, online sales, and exchange of information are thoroughly employed. With the dynamic development in the number of Internet-accessible services, Internet information security must be regularly maintained and adequate protection against cyberattacks is required. Use of traditional technology, such as a firewall, to repel attacks. Consequently, an IDS is typically set up to enhance the network security of businesses and other organizations [1]. A firewall is a passive system of manual protection, whereas an IDS system is an active system of automated protection.

An IDS (Intrusion Detection System), is the technique for detecting and tracking intrusive activities and reporting on any security breaches in an IT infrastructure or a network, along with analyzing evidence of potential events, such as unpredictability or imminent threats of breaching computer network security designs, acceptable implemented policies, or obsolete safety provisions. The two primary types of intrusion detection systems IDS are based on both signatures and abnormalities. The first method utilizes a repository that contains known malicious activity signatures along with generating an alert if communication over the network fits a specific signature, while the second one works with an approach for standardizing system operations and raises alarms for evident abnormalities. Although IDS have demonstrated their ability to identify multiple attacks on the network and have evolved into exhaustive protection in a broad network, it is of extreme concern that it triggers an excessive amount of alarms, almost all of which are false positives. This diminishes IDS's efficacy. Consequently, the overwhelming extent of improperly categorized warnings is a key issue influencing the operation of IDS [3]. False positive alerts are those that are triggered when an IDS erroneously identifies a benign action as harmful.

In the past couple of years, the proportion of erroneous alerts has been utilized to measure the effectiveness of Systems to identify intrusions; consequently, they play an essential part in reducing the overall result of these detection systems. IDSs are the most common approach for detecting such kind of behavior. IDSs include Intrusion Detection Using Signatures, Systems for detecting intrusions based on anomalies, and hybrid model Systems. Figure 1. Illustrates the various types of IDS.

Intrusion Detection Using Signatures employs algorithmic pattern-matching tactics to detect intrusions and preserve the signatures of harmful actions in a knowledge base. In the interim, AIDSs attempt to comprehend the typical conduct of the actions and designate comparable to the suspect. This form of system exists, a signature database is not required, and the system is capable of identifying Never-before-known zero-day vulnerabilities assaults observed before. The combination of SIDS and AIDS is used to aggregate hybrid systems to increase the prevalence of known malevolent actions while decreasing the percentage of erroneous positives for zero-day vulnerability assaults. [2].

Even though Signature-based IDS generates fewer false positive alerts than Anomaly-based IDS, false positives cannot be avoided. All of these explanations motivate researchers to develop strategies that can proportionally reduce false positives. Specifically, numerous technologies, encompassing machine learning algorithms as well as control visual content, and an intelligent false alarm filter, have been considered.

There are numerous kinds of detection of breaches strategies, but their accuracy is still a concern, accuracy is contingent on the frequency of detecting and erroneous alerts. To decrease a lot of erroneous positives and increase the frequency of detection, it is necessary to address the issue of precision. These research efforts were prompted by this concept. In the literature, the most common ML/DNN classifiers [4] are "LR, DT, KNN, Naive Bayes, Multinomial Naïve Bayes (MNB), Bernoulli Naïve Bayes (BNB), XG-Boost Classifier, Ada-Boost Classifier, RF, SVM, RC, Ridge, Passive-Aggressive (PA) classifier, ANN besides Perceptron (PPN), Back-propagation neural network (BPN) and stochastic-gradient descent (SGD)".

This paper intends to investigate the most advantageous techniques for eliminating erroneous alarms intrusion detection systems (IDSs) by analyzing the performance, of the majority prevalent algorithms for machine learning found in the literature on intrusion detection systems. In addition, matrices comparing the efficacy of the native classification algorithms have to be assessed analytically through precisely defined parameters that include "accuracy, precision, recall, and f-measure" on the openly accessible KDD Cup-99 dataset. The analysis will aid future researchers in gathering information on all proposed methods from a theoretical perspective and can serve as a fast reference for those unfamiliar with the topic.

In addition to firewalls, security administrators commonly employ password security tools, encryption methods, and access control to protect the network. However, these measures are insufficient to safeguard the system. As depicted in Figure 2, Numerous administrators prefer IDS for identifying malevolent keeping track of website traffic on network devices.

**Classification of an IDS (Intrusion Detection System)**

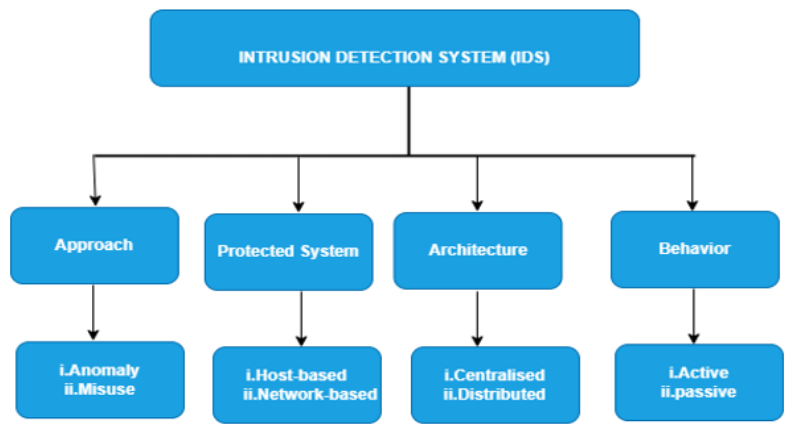

Figure 1: Types of IDS

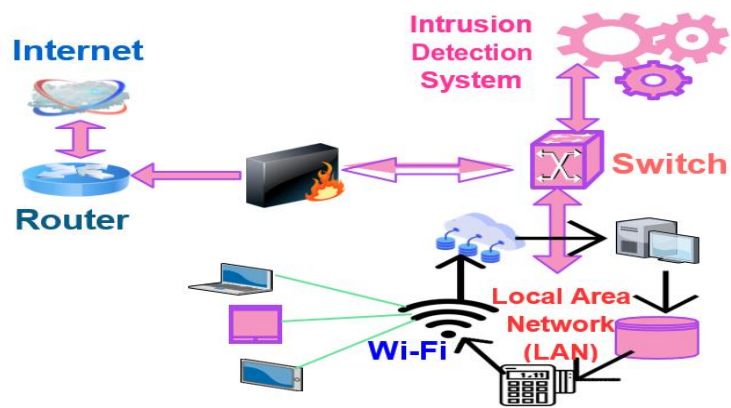

Figure 2: Area networks with local connectivity and IDS

## 2. INTRODUCTION TO THE PREFERRED STANDARD MODEL FOR IDS

In an endeavor to reduce the number of false positive alarm rates, we have recommended, as depicted in Figure 3, a five-step generic model for detecting Intrusion Detection Systems. Gathering raw datasets, preparatory data processing, extraction of attributes, and feature selection and detection are the fundamental steps.

**Data Pre-processing –** This entails structuring the collected data instances coming from the networking infrastructure necessities, and this can be input instantaneously within the ML algorithm. In this phase, also implemented are the techniques of separating features and selecting features.

**Training** - A procedure for learning through the machine is used to characterize the layouts of different categories of information to construct a pertinent computational simulation.

**Detection**- Subsequently the computational simulation has been generated, and data on traffic is going to be utilized as system input and compared to the computational simulation. If the data-driven sequence matches that of a known hazard, an alarm will sound. Classification, regression, and clustering are three typical problems that machine learning techniques can typically address. The detection of intrusions is a prevalent classifying dilemma. Consequently, labeled data collection for training is typically necessary for simulating computation. Several machine learning strategies have recently been implemented into security measures issues.

**Feature Selection**- This is a plan of action used to select and elucidate the relationship between significant data features. It facilitates the simplification of models and the reduction of the time required for the purpose to accomplish different outcomes via training and testing.

**Feature Extraction**- This phase is utilized if an amount of data is large and complex to compute. It is also used to transform readily accessible data in its original form into relatively easy data and to make use of the originally chosen information. The component for evaluating models incorporates the process of validating the validity embedded into the classifier based on cutting-edge metrics such as "accuracy, error rate, precision, recall, and F-1 scores".

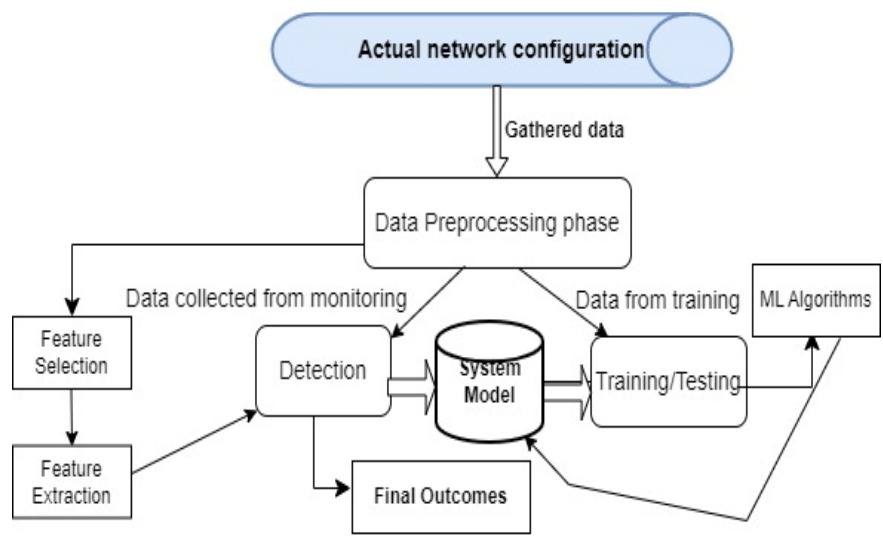

**Figure 3: Methodological parameters for constructing the preferred model for IDS**

## 2.1 System Architectural Design of IDS

As a result of, the fact that the recommended architecture is a hybrid of host and network-based Systems that detect intrusions, it is commonly known as a "Hybrid Intrusion Detection System (HIDS)". In the following illustration, one specific intrusion detection system captures packets and sends a call to a monitor agent, which then transmits the packets to a code-matching process. The code-matching process then checks attack parameters against the database, which has already resisted and uses stored rules for identifying an attack. After completing this process, an alarm will be activated if an attack is detected in the intercepted packet; otherwise, the alarm will be deactivated and this process will continue until the logical IDS system design is implemented. Figure 4. Depicts the system architecture.

Advances in Mechanical, Civil, Computer Engineering in respect Public Health and Safety

**DOI 10.17605/OSF.IO/WX6CS**

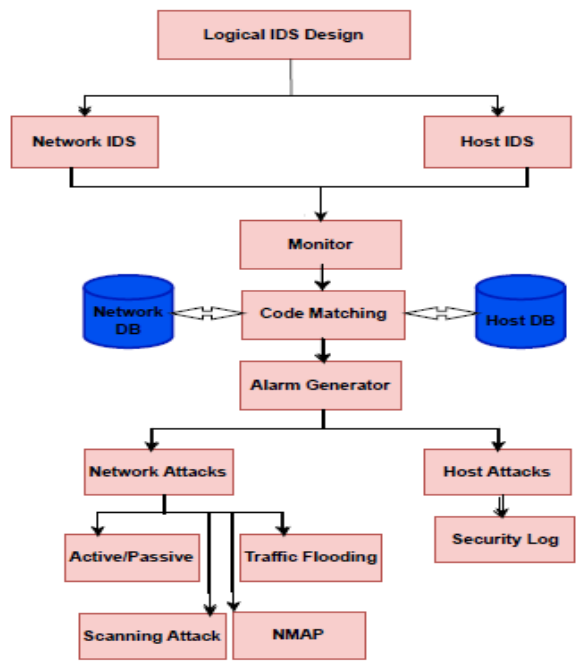

**Figure 4: Conceptual Layout of IDS**

## 2.2 Schematic Representation of Intrusion Detection System

The Figure 5. Illustrates an intrusion detection system block design. It includes the following components like Area Log File, Network analyzer, and Win Dump captures the headers of packets of data originating from the internet or a local area network. The data captured by Win Dump is saved to a file. The name of this file refers to a log file. Unit of data layout machine, data collected in a log file is categorized based on elements in the packet header. Using specific fields or predefined values for these fields, the protocols utilized by various packets are identified. Database Records contain distinct tables for various protocols, including TCP/IP, UDP, ICMP, and ARP. For each protocol, one table exists. Each table contains attributes pertinent to its respective protocol. The database stores layout data. Exploit Tracking Block, this technique of misuse detection is used to detect known attacks. Numerous computer attacks have a distinct signature. These signatures are suitable for tracking specific breaches. We compare the captured data packet header against a set of predefined criteria. If the pattern matches, the intrusion detection system classifies it as an intrusion and notifies the administrator. Similar to the log database, the attack database also contains tables for various protocols. The entries designated as attacks from the log database are stored in the attack database. This database can be referred to in the future for deriving conclusions or as a table displaying past system attacks.

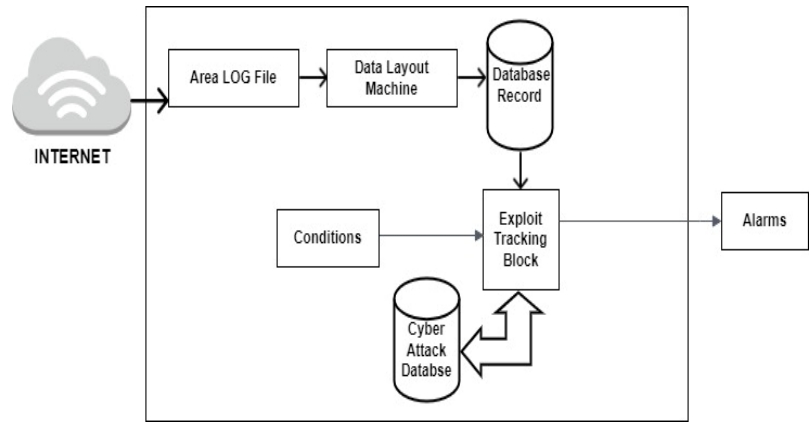

Figure 5: Schematic Representation of IDS

## 2.3 Literature Review

Different ML network-based detection systems known as Network Intrusion Systems (NIDSs) are being built recently to defend in opposition to malevolent attacks, according to the authors of [6]. This article recommends a literary work that is multiple stages optimized NIDS according to ML structure which minimizes the amount of computation required without compromising tracking efficacy, as well as the influence of excessive sampling approaches on the training sample number of prototypes and the minimum training sample number is examined and that is suitable. In addition, various ML hyper-factors optimization procedures are investigated to enhance the efficacy of the NIDS. The two current IDS datasets, CICIDS 2017 and UNSW-NB 2015 are utilized for assessing the usefulness of the suggested framework [6]. The findings of the experimental research demonstrate that the recommended paradigm substantially decreases the essential dimensions of samples for training (74% fit) and feature set size (50% fit). Furthermore, tuning parameter value optimization boosts the model's efficiency by finding accuracy scores above 99% among the two datasets, surpassing recent by 1-2% of literature better accuracy and the degree of precision and 1-2% fewer false positives.

The authors of [7] state that scholars suggested several ways of discerning suspicious activity to mitigate the impact of threats; moreover, up-to-date mechanisms frequently fail to adjust to constant change designs, linked hazards, and zero-day exploits. This script plans to address the flaws and drawbacks of existing datasets, their impact on Network Intrusion Detection System (NIDS) construction, along with the expanding quantity of refined hazards. This article concludes by employing a couple of essential facets of knowledge for researchers; an investigation of prevalent datasets, considering their implementation and influence on the advancement of the previous era's systems for recognizing Security breaches, and a categorization of network hazards and the auxiliary equipment to combat such assaults. The greatest aspect of the paper is merely the most recent IDS research comprises 33.3% of which of our hazard classification. The latest datasets reveal a distinct absence of actual network hazards, assault demonstration, along with a substantial proportion of criticized hazards, that limits the detecting efficiency of contemporary machine learning IDS techniques. This manuscript's novel integration of categorization and dataset investigation seeks to increase the development of datasets and the gathering of data from the current research globe.

Using the repeated existence of network security actions, the authors of [8] state that the Intrusion Detection System (IDS) is designed to produce records of alerts and logs when observing a network environment where numerous trace and alert records are stored unnecessarily, resulting in an unlimited burden on server storage and security workforces. Researchers have always been motivated by the question of how to diminish superfluous evidence of network detection of intrusion alerts. Using the whale optimization algorithm, the authors of this paper propose a method for handling a vast number of superfluous alarms. Based on the hierarchical clustering of alarms, they are incorporating the whale optimization technique within the method of alert generation. The UNSW-NB15 dataset is examined to validate the algorithm's probability; in comparison with earlier alarm clustering approaches, an alarm clustering approach developed using the whale optimization technique can produce a clustering of higher quality in less time. The outcomes indicate that the suggested technique could effectively diminish superfluous alarms and decrease the workload of IDS and personnel.

The development of systems for intrusion detection employs deep learning features in light of their technological advantages, including elevated correctness, computerization, and scalability, which enhance actual network intrusion detection systems, according to the authors of [9]. They proposed a Unique and adaptable resilient NIDS that employs a recurrent neural network (RNN) featuring a multi-classifier to produce a real-time detection model. In the outcomes of experiments, the system identifies network assaults with pinpoint accuracy and real-time simulation enhancement at a high rate, while demonstrating robustness under attack.

According to the authors of [10], a combination of techniques provides improvements in efficacy. They suggested a Double-Layer Hybrid Approach (DLHA) intended to deal with the aforesaid issue. They evaluated performance by applying the NSL-KDD dataset. The outcomes of the experiments indicate that DLHA surpasses extant IDS methods. By achieving detection rates of 96.67 percent and one hundred percent from R2L and U2R, respectively, DLHA excels at detecting rare attacks.

**Table 1: Comparative Summary of Related Works**

| Reference | Dataset | Classifiers Used | Assessment Matrix | Findings |
|---|---|---|---|---|
| S.Zwane, P.Tarwireyi et al. [11] | UNSW-NB15 | DT (J48), AdaBoost, RF, Bootstrap Aggregation. Multi-Layer Perceptron, Bayesian Network, SVM | TPR, FPR, ROC Curve, Test-time, and Build-time. | Collective Learning surpassed individual Ensemble-based models are typically sluggish in terms of model construction and evaluation. |
| M. Injadat, A. Moubayed et al. [6] | CICIDS 2017 and UNSW-NB 2015 | Bayesian Optimization implementing the Tree Parzen Estimator (BO-TPE-RF) and KNN | Precision, Accuracy, Recall/ TPR, FAR/FPR (false alarm/positive rate) | Concerning accuracy, Information Acquisition Based Selection of Features (IGBFS) outperformed correlation-based feature selection. |
| H. Hindy et al. [7] | KDD-99 | k-means versus SVM | TP, TN, FP, FN | It demonstrates that the latest IDS analysis only encompasses approximately 33.3% of the vulnerabilities prevalent across the taxonomies and that ML is utilized by 97.25 % of the analyzed IDS. |

Advances in Mechanical, Civil, Computer Engineering in respect Public Health and Safety

**DOI 10.17605/OSF.IO/WX6CS**

| Wang,L.;Gu, L.; Tang, Y. [8] | UNSW-NB15 | WOA (Whale Optimization Algorithm), hierarchical alarm clustering algorithm. | Accuracy Precisions, Recall | WOA has been used to address the issue of numerous superfluous alarms caused by an IDS to detect spurious alarms. |
|---|---|---|---|---|
| K. Yu, K. Nguyen et al. [9] | NSL-KDD 99, UNSW-NB15,Kyoto 2006,CIDDS | Multi-Classifiers like LR, DT, KNN, RF, Multilayer Perceptron, Gaussian_NB, and Gradient Boosting. | Based on the ROC (Receiver operating characteristic) curve, TPR, FPR, and Area under the curve (AUC) are utilized to characterize the performance of the proposed system. | This system uses RNN in conjunction with a multi-classifier and random system parameters to create a NIDS that is both adaptable and resilient. |
| T. Wisanwanichthan et.al. [10] | NSL-KDD | Naïve Bayes, SVM | Accuracy, F1_score, Precision, Detection rate (Recall), and FAR | The results of the experiments indicate that DLHA outperforms existing IDS methods. DLHA excels at detecting rare incidents with detection rates of 96.67 percent and one hundred percent from R2L and U2R, respectively. |
| R.Elhefnawy, H. Abounaser, et al. [12] | KDDCUP99, and UNSW-NB15 | FLS (Fuzzy Logic System) CLASSIFIER, SVM | Accuracy, Precision, Recall, FAR, F-Score | Hybrid Nested Genetic-Fuzzy Algorithm (HNGFA) outperforms other techniques for all types of assaults with a significant degree of FAR in a variety of dataset configurations. |
| Mahfouz, A. M., Venugopal, et al. [15] | NSL-KDD | Naïve Bayes, LR, Multilayer Perceptron, SVM, KNN, DT. | Accuracy, TPR, FPR, Precision, Recall, F-Measure, ROC Area | J48 surpasses other classifiers in terms of proficiency during the first phase, whereas IBK (Instance-Based KNN) executes faster throughout the second phase. |
| G. Kocher and G. Kumar, [16] | UNSW-NB15 | KNN, SGD, DT,RF,LR, Naïve Bayes | Accuracy, Mean Squared Error Precision, Recall, F1-Score, TPR, FPR. | RF classifier performs better than other classifiers. |
| Pai, Vasudeva & Bhat et al. [17] | NSL-KDD | RF, Naive Bayes, SVM, J48, LR, and DT | Accuracy, Precision, Recall, ROC, F1-Score | RF classifier performs better than other classifiers. |
| O. Almomani, M. A. Almaiah et al. [18] | UNSW-NB15 | LR, Multinomial Naive Bayesian, Gaussian Naive Bayesian, Bernoulli Naive Bayesian, KNN, DT, Adaptive Boosting, RF, Multilayer Perceptron, and Gradient Boosting | accuracy, precision, F-measure | Random Forest classifier boasts the greatest Accuracy, Precision, and F-measure for an IDS. |
| Othman et al. [19] | KDDCUP99 | SVM | AUROC (Area Under Curve), AUPR (Area Under | The paper's findings demonstrate that an automated FFNN (Feed Forward Neural Network) surpasses all other algorithms. |

|  |  |  | Precision-Recall Curve ) |  |
|---|---|---|---|---|
| Hasan, M., Islam et al. [20] | Kaggle | RF, ANN, LR, SVM. | Accuracy, precision, recall, f1 score, ROC. | The test demonstrates that RF is an effective approach for IDS with a detection rate of 99.4% in IoT. |
| R. Kumar Singh Gautam et al. [21] | KDD-99 | Naive Bayes, PART (partial decision tree algorithm), Adaptive Boost, and ensemble methods. | Precision, Recall, Accuracy. | The results of the study indicate that the ensemble methodology employing bootstrap approaches outperforms the other classifier. |

## 3. INTRUSION DETECTION SYSTEMS WITH ML CLASSIFIERS

### 3.1 Logistic Regression

In the approach of the logistic regression classification paradigm, the logistic regression classifier is typically employed to anticipate a categorical response. It can address problems related to both binary data and multiple classifications. One can estimate the likelihood of a specific event occurring through data that fits the logistic function. The outcome range for this feature is 0 to 1. The significance of 0.5 represents the boundary within classes 1 and 0. The result is specified as category 1 exceeds 0.5, along with the outcome specified as class 0 which falls below 0.5. Logistic regression is an algorithm for classifying observations into discrete classes [14]. As opposed to logistic regression, which modifies the result using the logistic sigmoid function to produce a value for the probability that can be applied to two or more distinct classifications. $F(x) = 1 / 1+e^{-x}$ describes the sigmoid function.

x represents the function's input parameter, while F(x) returns a value between 0 and 1; e is the natural log base.

### 3.2 Naive Bayes Classifier

Naïve Bayes is a straightforward classification of possibility based on the theorem of Bayes in which every attribute or parameter is taken to be unchanged from one another. Conditional probability maintains the constraint with a distinct relationship among the features. There exist both variants of the nave Bayes classifier; both the Multivariate Bernoulli model (B_NB) and the Multinomial Bernoulli model (M_NB) are included. The multivariate Bernoulli naive Bayes model operates exclusively with information in binary format. Thomas Bayes, who was a British scientist, proposed Bayes' theorem as an approach for anticipating future possibilities based on past data [22]. The Bayes theorem is represented by the following formula (1)

$$P(E|F) = \frac{P(F|E) \cdot P(E)}{P(F)} \quad (1)$$

Where: F: The facts along with unclassified groups

E: The F facts statement belongs to a particular group

P (E|F): The likelihood of an assumption E is dependent on the presence of state F

P (E): Assumption likelihood E

P (F|E): F likelihood depending on states

P (F): F likelihood

### 3.3 K-Nearest Neighbor Classifier (KNN)

KNN classification technique was similar to cases that are also known as the "lazy learner" due to the absence of a learning period. It only generates results when they are requested. A classifier based on KNN is utilized for the classification of program behavior as typical or invasive [23]. KNN is a supervised classifier that predicts the outcome of the intended parameter via locating the k-nearest neighbors using distance by Euclidean measurement. It is a non-parametric method of classification that does not make any inferences regarding the data being classified [24].

**KNN Algorithm:**

Step-1: Input the training and testing datasets.

Step-2: Select k as the value of the neighbors

Step-3: In every instance of the data specimen, find out the difference between the specimen and its neighbors.

- Save the interspaces and arrange them ascendingly.
- Acknowledge the initial k responses.
- Classify the most recent dataset according to the preponderance classes included in the neighboring position.

Step 4: Document the final precision.

Step 5: Our prototype is all set

### 3.4 Decision Tree Classifier

The Decision tree algorithm is one of which is most widespread classification techniques. The decision tree is an example of a tree-shaped graph. It assigns classification based on the principles applied from the trunk to the leaves of the tree. The internal nodes are tests, the branch corresponds to the test result, and the leaf nodes are classified.

**Decision Tree algorithm**

1. Select any attribute from the data set.
2. Calculate the importance of each attribute when dividing the data.
3. Separate the data based on the greatest attribute's value.
4. Again, return to step 1.

### 3.5 Extreme Gradient Boosting (XGBOOST) Classifier

Ensemble learning is based on the assumption that the error of one model's classifier can be mitigated by other classifiers when the model involves multiple models, which provide superior performance than a single model. Boosting is a technique that recurrently combines weak classifiers (barely better than random) into a more accurate model. XGBoost is a form of boosting technology that employs trees as base learners. It is a scalable tree-growing system. XGBoost is developed upon a framework for gradient boosting. Gradient boosting is a machine learning method used to solve classification, regression, and clustering issues. When making predictions, it optimizes the model as depicted in Figure 6.

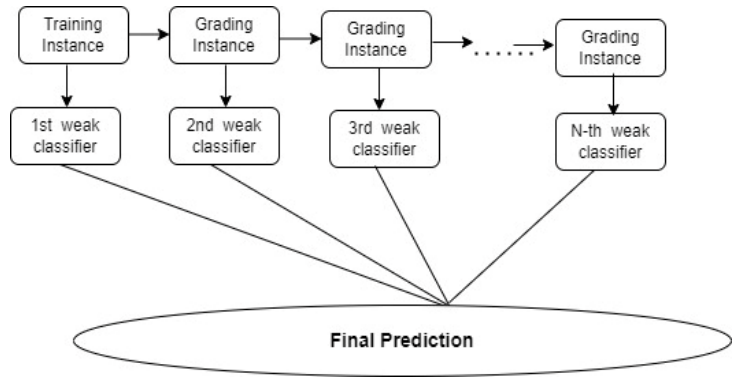

Figure 6: Extreme Gradient Boosting Classifier

### 3.6 AdaBoost Classifier

Based on the features of the AdaBoost technique and the network detection of intrusion issues, our approach's framework consists of the four modules depicted in Figure 7, extraction of features, data labeling, designing the weak classifiers, and building the robust classifier. AdaBoost is one of the most well-known algorithms for constructing an ensemble classifier from weak member classifiers. The AdaBoost algorithm produces a robust classifier that is a mixture of several weak classifiers. AdaBoost identifies a combination of weak classifiers with weight adjustments through a process of iteration, while the original training data set remains unmodified. The diversity of feeble classifiers is one of the reasons which explains the AdaBoost algorithm achieves favorable outcomes.

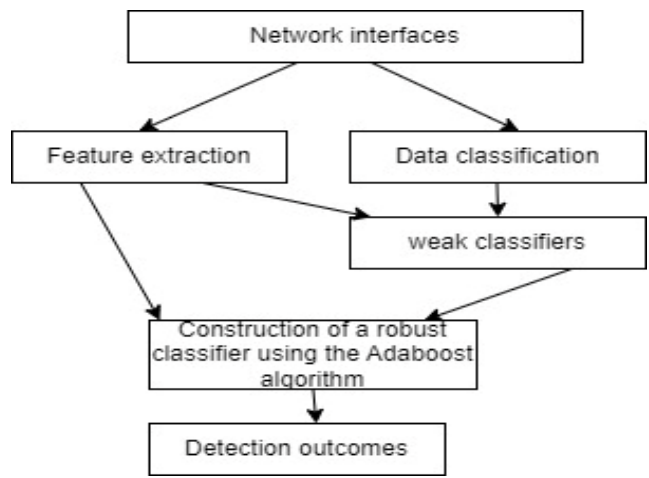

Figure 7: Structure of an AdaBoost algorithm

### 3.7 Random Forest Classifier

Based on the characteristics of the AdaBoost algorithm, the random forests principle is a combination of classification and regression methods, which is among the majority of efficient data mining techniques. The random forest algorithm has been widely implemented in a variety of applications. For instance, it has been utilized for the prediction and estimation of probability. Random forest constructs multiple decision trees and combines them to produce a more precise and stable forecast. There are numerous benefits of random forests. Individual decision trees have the propensity to overfit the training data, but random forests can mitigate this problem by aggregating the prediction results from multiple trees. This makes random forests more accurate predictors than a single decision tree depicted in Figure 8.

**Random Forest Algorithm:**

**Step-1:** The procedure draws specimens arbitrarily from the supplied data set.

**Step-2:** The procedure involves the development of a decision tree for every specimen. Next, it will obtain each constructed decision tree to yield an estimate.

**Step 3:** Involve casting their votes for each estimated outcome. It will utilize mode regarding classification issues and mean for issues related to regression.

**Step-4:** The algorithm is going to choose the prediction with the majority of polling as its outcome.

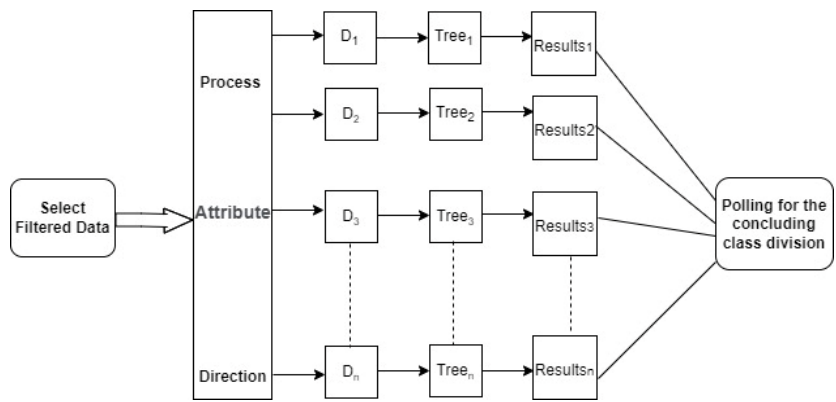

**Figure 8:** The architectural design of the Random Forest for IDS

### 3.8 Support Vector Machine (SVM)

Systems based on SVM for intrusion detection and feature selection system. SVM, also known as an SVM is a machine learning approach that is founded on the concept of the supervised machine learning model. The support vector machine (SVM) uses the idea of statistical learning to classify records by determining a collection of support vectors, which are members of the labeled data used for training samples. SVM is a classification method that is capable of classifying a combination of nonlinear as well as linear data. The fundamental creativity underlying classification by SVM is that it initially non-linearly maps the first training session and incorporates data split into a far greater dimension, say n, so that the data in the higher dimensionality can be readily differentiated by (n-1) dimensionality decision surfaces referred to as hyperplanes. Classification by SVM recognizes the most basic hyperplane with the greatest margins from the support vectors. The primary objective of a support vector machine is to process of establishing the most appropriate hyperplane for the classification of novel data elements.

### 3.9 Artificial Neural Network (ANN)

ANN is a basic record of data handling nonlinear model admire the framework of the human brain cognitive system, and it will acquire knowledge from the exhaustive data on training to execute operations such as classification, estimation or predicted outcomes, the formulation of decisions, graphical representation, in addition to others as well. It is comprised of a set of nodes, commonly referred to as neurons, that constitute the fundamental unit of dealing with information in ANN. About the dilemma assertion, these neurons are arranged into three separate layers; the input layer, the output layer, and the hidden layer. In the framework of classification of false positive threats, the number of threats determines the total number of neuronal cells in the input or first layer, whereas the filter of network classification categories influences the percentage that comprises neuronal cells as part of each output or final layer. ANN will include at least a single input or first layer along with a single output or final layer but will have several hidden layers based on the chosen disadvantage. Every connection regarding the source of input to the result of the output layer with the help of layers that are hidden is assigned weights that illustrate the degree of dependence between nodes. When neurons obtain weighted data, an established activation function analyses the weighted sum. To identify the appropriate neuron in the results of the output layer, the activation function's output value is provided to every neuron through its input layer. Rectifier linear unit (ReLU) features consist of Binary step, sigmoid function TanH, and Softmax. Are some commonly utilized instances of activation functions by utilizing extra hidden layers, ANN might become more adaptable and efficient. PPN, SGD, and BPN consist of all three predominant equipped with neural networks that operate the algorithms that are widely utilized for reducing false alarm rates.

## 4. RIDGE CLASSIFIER (RIDGE)

The Ridge method of classification relies on the domain postulation that instances of specific categories reside on a portion of space that is linear and that the latest class can be administered screening of the sample defined in the form of a continuous arrangement of sample training evaluation for the pertinent category [27]. Ridge regression is an approach for analyzing multicollinear data that has been used for algorithm-optimized performance. This approach pulls out L2 regularization. When the issue of multicollinearity occurs, the least-squares are random and variances are at large, leading to predicted values that are significantly lower than the actual value L1 and L2 may decrease the complexity of the model and prevent overfitting caused by simple linear regression.

### 4.1 Passive Aggressive (PA) Classifier

Classifiers who employ passive aggressions are parts of the huge-scalability method of learning technique category [28]. It is one of them the few online-learning algorithms known. In online machine learning algorithms, the input data arrives periodically and the machine learning model gets modified continually, as opposed to sequential learning, which utilizes the whole training set at once. The concept of operation of such a classification style is similar to the perception network encoder; despite this, no learning rate is required. Still, it contains the regularisation parameter C. Here C is the regularisation parameter, and it indicates the penalty that the model will impose due to a wrong prediction.

### 4.2 Rocchio Classifier (RC)

The Rocchio algorithms for classification rely on the well-established response to relevance theory in the area of data retrieval. It utilizes the characteristic's measure of midpoint and proximity estimates between false alarm rates during the instruction and evaluation phases of the prototype creation and application, accordingly. During the instruction phase, the Rocchio classifier calculates the midpoint for every classification-related online web attack and recognizes the centroid as the representative for each class. During the evaluation phase, the Rocchio classifier estimates its Euclidean measurement from the midpoint of every classification to determine the grouping label of an undetected false alarm rate.

## 5. EFFICIENCY MEASUREMENT OF AN IDS

The efficiency of the Intrusion Detection System is crucial for enhancing computer security. It provides service developers with the data and inferences necessary to enhance their IDS and informs consumers of the IDS's strengths and weaknesses. The successful operation of an IDS is measured based on its ability to operate accurately and classify events as attacks or normal behavior based on its predictive capabilities. According to the actual environment of a particular event and the estimation from the IDS, Table 2. Presents four probable conclusions.

**Table 2: Calculating IDS Efficiency Components**

| Real Class | Expected Class | |
|---|---|---|
|  | **Standard** | **Assault** |
| Standard | True Negative($T_{-ve}$) | False Positive($F_{+ve}$) |
| Assault | False Negative($F_{-ve}$) | True Positive($T_{+ve}$) |

**True Negative (T-ve):** The number of accurate predictions that an instance does not belong to a specific class.

**False Positive (F+ve):** The number of erroneous predictions that an instance belongs to the same class when it is a class of a distinct type.

**False Negative (F-ve):** The number of erroneous predictions that a situation pertains to a different class when it belongs to the same category.

**True Positive (T+ve):** The number of accurate predictions that an instance falls into the identical category.

The rate of false positives (FPR), also referred to as the rate of false alarms (FAR), is the rate at which typical information is erroneously identified as an assault. An enormous FPR will significantly degrade the efficacy of the IDS, while an essential False Positive Rate (FNR) will render the system susceptible to commands. To optimize IDS acts, the prevalence for FP and FN needs to be decreased whereas efficiency is increased. All suggested methods regarding diminishing erroneous positives are inadequate as a result of the fact that diminishing false positives alone is insufficient. Therefore, it is essential to implement techniques that reduce the number of false positives while maintaining or improving accuracy.

Existing ML-based intrusion detection system (IDS) evaluation metrics are diverse; however, the objective of this study is to maximize the total volume of occurrences in which the dataset used for testing estimations is exact as well. Accuracy is the most important metric to be considered.

$$\text{Accuracy} = \frac{T(+ve)+T(-ve)}{T(+ve)+T(-ve)+F(+ve)+F(-ve)} \quad (1)$$

$$\text{Precision} = \frac{T(+ve)}{T(+ve)+F(-ve)} \quad (2)$$

$$\text{Recall} = \frac{T(+ve)}{T(+ve)+F(+ve)} \quad (3)$$

$$\text{F1 Score} = \frac{2(\text{recall} * \text{precision})}{\text{recall} + \text{precision}} \tag{4}$$

$$\text{False Positive Rate (FPR)} = \frac{F(+ve)}{F(+ve) + T(-ve)} \tag{5}$$

$$\text{False Negative Rate (FNR)} = \frac{F(-ve)}{F(-ve) + T(+ve)} \tag{6}$$

$$\text{True Positive Rate (TPR)} = \frac{T(+ve)}{T(+ve) + F(-ve)} \tag{7}$$

$$\text{True Negative Rate (TNR)} = \frac{T(-ve)}{T(-ve) + F(+ve)} \tag{8}$$

## 6. RESULTS AND DISCUSSION

### 6.1 Experimental Setup

Python Scikit-learn libraries are utilized to implement techniques of ML computations. From start to finish testing is performed using the cloud service, Google Colaboratory, which has a GPU with an embedded Tesla K20, 2496 CUDA cores, 16GB of RAM, and 50 GB of disc capacity. Experiments presented in this paper were conducted on a DESKTOP-UFN62J4 running Windows 11 and equipped with the following processor i.e. 11th Generation Intel Core i3-1115G4 @ 3GHz 3 GHz.

### 6.2 Dataset Description

IDSs can be created either based on signatures or anomalies. A dataset should be used to train normal and anomalous requests to detect system anomalies. Researchers can utilize either a publicly available dataset or their datasets. In the following subsections, the content and properties of the most popular datasets are compared and contrasted. Applying the KDD Cup-99 data set, experiments were conducted. Over 70% of the KDD cup-99 dataset was put to use to train and 30% was used for testing using 10 Fold cross-validation in this study. Afterward this experiment, the data set had been turned into a resource for Intrusion Detection literature, cited across many academic papers. Since the publication of the KDD-'99 dataset in 1999, it has been the most widely used data for analyzing IDSs. This dataset is comprised of nearly 805050 unique connections and 41 distinct characteristics. The simulated assaults were broadly classified as follows:

(i) DoS (Denial-of-Service); (ii) R2L (Remote-to-Local); (iii) U2R (User-to-Root); and

(iv) Probing are the four categories of KDD Cup-99 attacks.

**(i) Normal:** Non-attack data that is typical connections can be created by simulating normal user actions, which includes accessing files and browsing on the World Wide Web.

**(ii) Denial of Service (DoS) Attacks:** These kinds of assaults typically prevent users from receiving services by providing repetitive requests for connection to a server in breaches of the TCP/IP protocol's authentication framework. (e.g. Syn flood)

**(iii) Probe Attacks:** These kinds of assaults are used to locate specific data on a server or other machine. (e.g. port scanning)

**(iv) Remote to User (R2L) Attacks:** These are assaults carried out leading to unauthorized guest login or login as a different user. (e.g. guessing password)

**(v) User to Root (U2R) Attacks:** The points that follow are the assaults of a user who is permitted to access this mechanism but is not an administrator, by employing this approach of assault, a user can impersonate a computer programmer or an admin and conduct unauthorized operations. (e.g. various "buffer overflow" attacks).

**Table 3: Case counts for each attack category**

| Normal | 67343 |
|---|---|
| DoS | 45927 |
| Probe | 11656 |
| R2L | 995 |
| U2R | 52 |

Table 3. Displays the number of cases per attack category within the dataset

### 6.3 Results

This section examines the results of a comprehensive experiment performed on machine learning algorithms using KDD-based security detection data sets, consisting of KDD Cup-99. In the Scikit- learn ML library, all ML methods have been implemented. Consequently, 10 Fold cross-validation is used to train all ML algorithms. In 10 Fold cross-validation, the prototype trained of machine learning ML computational methods is conducted across tenth cycles. Within each repetition, the dataset's intrusions are evenly divided into ten parts, with each part selecting intrusions at random from the entire dataset. On average, nine out of ten sections are used for training, while the remaining portion is put to use in the evaluation process. After the tenth repetition, the mean and deviations from the mean of each efficiency indicator are calculated. All machine learning computational methods utilized the conventional hyper-parameter variables specified by the Scikit-learn ML library.

**Table 4: Performance of ML algorithms for KDD-based Network IDS dataset**

| Classification Algorithms | Performance Measure (Mean ± Deviation) | | | | |
|---|---|---|---|---|---|
| | Accuracy | Precision | Recall | F1-Score | Execution Time in Sec. |
| KNN | 0.9724±0.0551 | 0.9806±0.0311 | 0.9724±0.0551 | 0.9741±0.04990 | 4494.731sec |
| DT | 0.9713±0.0556 | 0.9809±0.3011 | 0.9713±0.0556 | 0.9731±0.0502 | 138.607sec |
| MNB | 0.9329±0.3869 | 0.9399±0.2870 | 0.9329±0.0.3868 | 0.9298±0.0389 | 5.229sec |
| BNB | 0.9473±0.0555 | 0.9594±0.0272 | 0.9473±0.0555 | 0.9487±0.0497 | 8.268sec |
| RF | 0.9714±0.0590 | 0.9811±0.0309 | 0.9714±0.0590 | 0.9733±0.0531 | 1479.658sec |
| SVM | 0.9808±0.0061 | 0.9815±0.0053 | 0.9808±0.0061 | 0.9808±0.0059 | 39600.860sec |
| PPN | 0.9101±0.1303 | 0.9539±0.0430 | 0.9101±0.1303 | 0.9157±0.1186 | 17.751sec |
| LR | 0.9667±0.0595 | 0.9777±0.0289 | 0.9667±0.0595 | 0.9689±0.0534 | 1445.899Sec |
| XGBoost | 0.9464±0.0827 | 0.9646±0.0325 | 0.9464±0.0827 | 0.9496±0.0736 | 1318.714Sec |
| AdaBoost | 0.9431±0.0598 | 0.9556±0.0288 | 09431±0.0598 | 0.9449±0.0530 | 120.88Sec |
| SGD | 0.9046±0.1255 | 0.9457±0.0389 | 0.9046±0.1255 | 0.9097±0.1136 | 27.357sec |
| Ridge | 0.9495±0.0395 | 0.9562±0.0241 | 0.9495±0.0395 | 0.9499±0.0359 | 64.057sec |
| RC | 0.9487±0.0388 | 0.9549±0.0251 | 0.9487±0.0388 | 0.9488±0.0350 | 5.107sec |
| PA | 0.9443±0.0650 | 0.9576±0.0278 | 0.9443±0.06000 | 0.9462±0.0532 | 21.236sec |
| BPN | 0.9704±0.5934 | 0.9810±0.0290 | 0.9704±0.0594 | 0.9725±0.0534 | 9555.755sec |

The mean and standard deviation of the classification performance are presented in Table 4.

And Figure 9. Compares and displays graphically the classification accuracy of the ML algorithms. Among the various classification algorithms for which we have obtained different levels of accuracy, The Classifier based on the support vector machine (SVM) concept demonstrates the highest classification effectiveness, with a score associated with 98.08%. In addition to KNN, the DT, RF, and BPN classifiers offer good performance when it comes to classification. When compared to the other classifiers, the SGD classifier demonstrates the worst performance for the KDD Cup-99 dataset.

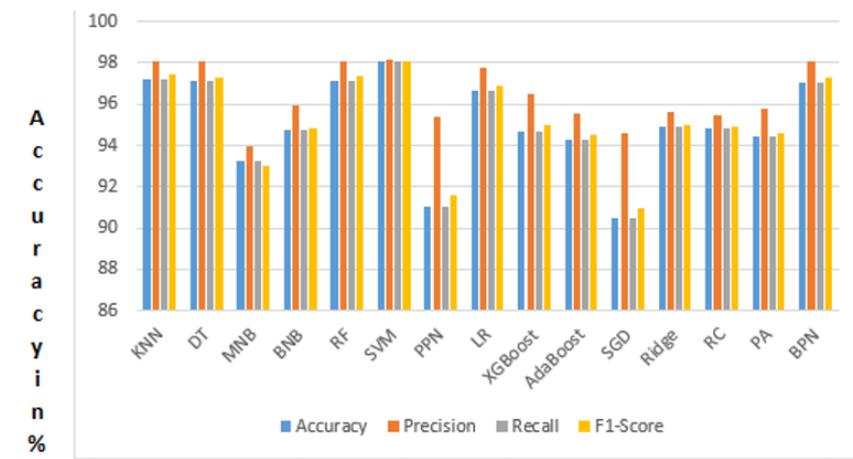

Figure 9. Performance Comparisons of ML Classification Algorithms in IDS

## 7. CONCLUSION AND FUTURE SCOPE

LR, DT, KNN, Naive Bayes, Bernoulli Naïve Bayes (BNB), Multinomial Naïve Bayes (MNB), XGBoost, AdaBoost, RF, SVM, Rocchio Classifier (RC), Ridge, PA Classifier, ANN along with Perceptron (PPN), SGD and BPN classifiers for an IDS have been evaluated in this present research. To evaluate these classifiers, the KDD CUP-99 dataset was used. Accuracy, Precision, Recall, and f1-measure are employed for evaluating the reliability of this ML classifier. Among the various algorithms for which we have obtained varying degrees of precision, in terms of accuracy Support Vector Machine gives the highest accuracy rate 98.08%. The graph demonstrates that, compared to other algorithms, the SVM algorithm used in this research yields greater accuracy. Therefore, these process steps are outlined extensively in the current research involved in the intrusion detection process, illustrate the computational facts of the cutting-edge supervised machine learning algorithms and explicitly determine how each algorithm used in ML is employed that are designed to serve as classifiers for the standards of KDD CUP-99 dataset performance. In comparison to the other methods for classification including SVM, DT, RF, KNN, and BPN produce the best outcomes on the provided dataset. However, SGD and PPN classifiers performed poorly across the selected data set when in contrast to different classification techniques. Meanwhile, the classification efficiency of other classifiers is average. Future objectives include enhancing the adaptability of these classifiers to large-scale datasets. Consequently, the implementation of prototypes under deep learning techniques, consisting of Multilayer Feed Forward Neural Networks (MFFNN), CNN, and RNN refer to convolutional and recurrent neural networks as well as ensemble deep learning models and Extreme Learning Machine (ELM) has become an unavoidable direction for future research.